\documentclass[longauth]{aa}       

\usepackage[skip=4pt]{caption}
\usepackage[utf8]{inputenc}
\usepackage{multirow}
\usepackage{natbib}
\usepackage{url}
\usepackage{xspace}
\usepackage{graphicx, pstricks}
\PassOptionsToPackage{dvipsnames}{xcolor} 
\usepackage[colorlinks=True,
            allcolors=blue]{hyperref}
\usepackage{cleveref}
\graphicspath{{./gfx/}}

\usepackage{catoptions}
\makeatletter

\def\Autoref#1{%
  \begingroup
  \edef\reserved@a{\cpttrimspaces{#1}}%
  \ifcsndefTF{r@#1}{%
    \xaftercsname{\expandafter\testreftype\@fourthoffive}
      {r@\reserved@a}.\\{#1}%
  }{%
    \ref{#1}%
  }%
  \endgroup
}
\def\testreftype#1.#2\\#3{%
  \ifcsndefTF{#1autorefname}{%
    \def\reserved@a##1##2\@nil{%
      \uppercase{\def\ref@name{##1}}%
      \csn@edef{#1autorefname}{\ref@name##2}%
      \autoref{#3}%
    }%
    \reserved@a#1\@nil
  }{%
    \autoref{#3}%
  }%
}
\makeatother

\newcommand{\gaia}{\textsl{Gaia}\xspace}
\newcommand{\astorb}{\texttt{astorb}\xspace}
\newcommand{\aladin}{\texttt{Aladin}\xspace}
\newcommand{\aladinlite}{\texttt{AladinLite}\xspace}

\newcommand{\numb}[1]{\textcolor{orange}{#1}\xspace}
\renewcommand{\numb}[1]{#1\xspace}

\newcommand{\mcmc}{\numb{2000}}

\DeclareUnicodeCharacter{0441}{C}

\begin{document} 
  \title{Potential asteroid discoveries by the ESA \gaia mission}
  \subtitle{Results from follow-up observations}
  \author{
     B.~Carry\inst{\ref{imcce},\ref{oca}}   \and
     W.~Thuillot\inst{\ref{imcce}}          \and
     F.~Spoto\inst{\ref{oca},\ref{mpc}}     \and
     P.~David\inst{\ref{imcce}}             \and
     J.~Berthier\inst{\ref{imcce}}          \and
     P.~Tanga\inst{\ref{oca}}               \and
     F.~Mignard\inst{\ref{oca}}             \and
     S.~Bouquillon\inst{\ref{syrte}}        \and
     R.~A.~Mendez\inst{\ref{santiago}}      \and
     J.-P.~Rivet\inst{\ref{oca}}            \and
     A.~Le~Van~Suu\inst{\ref{ohp}}          \and
     A.~Dell'Oro\inst{\ref{firenze}}        \and
     G.~Fedorets\inst{\ref{hel},\ref{qub}}  \and
     B.~Frezouls\inst{\ref{dpcc}}           \and
     M.~Granvik\inst{\ref{hel},\ref{ltu}}   \and
     J.~Guiraud\inst{\ref{dpcc}}            \and
     K.~Muinonen\inst{\ref{hel},\ref{fgi}}  \and
     C.~Panem\inst{\ref{dpcc}}              \and
     T.~Pauwels\inst{\ref{rob}}             \and
     W.~Roux\inst{\ref{dpcc}}               \and
     G.~Walmsley\inst{\ref{dpcc}}           \and
     J.-M.~Petit\inst{\ref{besac}}          \and
     L.~Abe\inst{\ref{oca}}                 \and
     V.~Ayvazian\inst{\ref{g1},\ref{g2}}    \and
     K.~Bailli{\'e}\inst{\ref{imcce}}       \and
     A.~Baransky\inst{\ref{kyiv}}           \and
     P.~Bendjoya\inst{\ref{oca}}            \and
     M.~Dennefeld\inst{\ref{iap}}           \and
     J.~Desmars\inst{\ref{imcce},\ref{ipsa}}\and
     S.~Eggl\inst{\ref{imcce},\ref{lsst}}   \and
     V.~Godunova\inst{\ref{icamer}}         \and
     D.~Hestroffer\inst{\ref{imcce}}        \and
     R.~Inasaridze\inst{\ref{g1},\ref{g2}}  \and
     V.~Kashuba\inst{\ref{odessa}}          \and
     Y.~N.~Krugly\inst{\ref{kharkiv}}       \and
     I.~E.~Molotov\inst{\ref{ras}}          \and
     V.~Robert\inst{\ref{imcce},\ref{ipsa}} \and
     A.~Simon\inst{\ref{kyiv2},\ref{kyiv3}} \and
     I.~Sokolov\inst{\ref{terskol}}         \and 
     D.~Souami\inst{\ref{lesia},\ref{naxys}}\and
     V.~Tarady\inst{\ref{icamer}}           \and
     F.~Taris\inst{\ref{syrte}}             \and
     V.~Troianskyi\inst{\ref{odessa}}       \and
     V.~Vasylenko\inst{\ref{kyiv2},\ref{kyiv3}} \and
     D.~Vernet\inst{\ref{oca}}  
}

  \institute{
  Institut de M{\'e}canique C{\'e}leste et de Calcul des {\'E}ph{\'e}m{\'e}rides IMCCE, Observatoire de Paris, Universit{\'e} PSL, CNRS, Sorbonne Universit{\'e}, Universit{\'e} de Lille, 77 av. Denfert Rochereau,75014 Paris, France\label{imcce}
  \and
  Universit{\'e} C{\^o}te d'Azur, Observatoire de la C{\^o}te d'Azur, CNRS, Laboratoire Lagrange, Boulevard de l'Observatoire, CS34229, 06304, Nice Cedex 4, France\label{oca}
  \email{benoit.carry@oca.eu}
  \and
  Harvard-Smithsonian Center for Astrophysics, 60 Garden St., MS 15, Cambridge, MA 02138, USA\label{mpc}
  \and
  SYRTE,Observatoire de Paris, PSL Research University, CNRS, Sorbonne Universit{\'e}, UPMC Univ. Paris 06, LNE, 61 avenue de l'Observatoire, 75014 Paris, France\label{syrte} 
  \and
  Departamento de Astronom{\'i}a, Facultad de Ciencias F{\'i}sicas y Matem{\'a}ticas, Universidad de Chile, Casilla 36-D, Santiago, Chile\label{santiago}
  \and
  Aix Marseille University, CNRS, Institut Pytheas-Observatoire Haute Provence, F-04870 St-Michel-l'Observatoire, France\label{ohp} 
  \and
  INAF - Osservatorio Astrofisico di Arcetri, Largo Enrico Fermi 5, 50125 Firenze, Italy\label{firenze}
  \and
  Department of Physics, Gustaf H{\"a}llstr{\"o}min katu 2, University of Helsinki, PO Box 64, 00014, Finland\label{hel}
  \and
  Astrophysics Research Centre, School of Mathematics and Physics, Queen’s University Belfast, Belfast BT7 1NN, UK\label{qub}
  \and
  CNES Centre Spatial de Toulouse, 18 avenue Edouard Belin, 31401 Toulouse Cedex 9, France\label{dpcc}
  \and
  Asteroid Engineering Lab, Onboard Space Systems, Lule\aa{} University of
  Technology, Box 848, S-981 28 Kiruna, Sweden\label{ltu} 
  \and
  Finnish Geospatial Research Institute, Geodeetinrinne 2, 02340, Masala, Finland\label{fgi}
  \and
  Royal Observatory of Belgium, Avenue Circulaire 3, B-1180 Bruxelles, Belgique\label{rob}
  \and
  Observatoire de Besan{\c c}on, UMR CNRS 6213, 41 bis avenue de l'Observatoire, 25000, Besan{\c c}on, France\label{besac}
  \and
  Kharadze Abastumani Astrophysical Observatory, Ilya State University, K.
  Cholokashvili Avenue 3/5, Tbilisi 0162, Georgia\label{g1}
  \and
  Samtskhe-Javakheti State University, Rustaveli Street 113, Akhaltsikhe 0080,
  Georgia\label{g2}
  \and
  Astronomical Observatory, Taras Shevсhenko National University of Kyiv, 3 Observatorna str., Kyiv, 04053, Ukraine\label{kyiv}
  \and
  Institut d'Astrophysique de Paris, Sorbonne Universit{\'e}, CNRS, UMR 7095, 98 bis bd Arago, 75014, Paris, France\label{iap} 
  \and
  Institut Polytechnique des Sciences Avanc{\'e}es IPSA, 63 bis Boulevard de Brandebourg, F-94200 Ivry-sur-Seine, France\label{ipsa}
  \and
  Vera C. Rubin Observatory/DIRAC Institute, Department of Astronomy, University of Washington, 15th Ave. NE, Seattle, WA 98195, USA\label{lsst} 
  \and
  ICAMER Observatory of NASU, 27 Acad. Zabolotnogo str., Kyiv, 03143, Ukraine\label{icamer}
  \and            
  Astronomical Observatory of Odessa I.I. Mechnikov National University, 1v Marazlievska str., Odessa, 65014, Ukraine\label{odessa}
  \and
  Institute of Astronomy, V.N. Karazin Kharkiv National University, Sumska Str. 35, Kharkiv, 61022, Ukraine\label{kharkiv}
  \and
  Keldysh Institute of Applied Mathematics, RAS, Miusskaya sq. 4, Moscow, 125047, Russia\label{ras}
  \and
  Astronomy and Space Physics Department, Taras Shevchenko National University of Kyiv,  60 Volodymyrska str., Kyiv, 01601, Ukraine\label{kyiv2}
  \and
  National Center ``Junior Academy of Sciences of Ukraine'',  38-44, Degtyarivska St., Kyiv, 04119 , Ukraine\label{kyiv3}
  \and
  Terskol Branch of INASAN RAN, 48 Pyatnitskaya str., Moscow, 119017, Russia\label{terskol}
  \and
  LESIA, Observatoire de Paris, Sorbonne Universit{\'e}, Universit{\'e} PSL, CNRS, Univ. Paris Diderot, Sorbonne Paris Cit{\'e}, 5 place Jules Janssen, F-92195 Meudon, France\label{lesia}
  \and
  naXys, University of Namur, Rempart de la Vierge, Namur 5000, Belgium\label{naxys}
  \and
  Astronomical Observatory Institute, Faculty of Physics, A. Mickiewicz University, Słoneczna 36, 60-286 Pozna{\'n}, Poland\label{amu}
}

  \date{Received May XX, 2020; accepted June XX, 2020}

  \abstract
   {Since July 2014, the \gaia mission of the European Space Agency has been surveying the
   entire sky down to magnitude 20.7 in the visible.
   In addition to the millions of daily observations of stars, thousands of
   Solar System Objects (SSOs) are observed. By comparing their positions,
   as measured by \gaia, to those of known objects, a daily processing pipeline filters
   known objects from potential discoveries.
   However, owing to \gaia's specific observing mode,
   which follows a pre-determined scanning law
   designed for stars as \textsl{fixed} objects on the celestial sphere,
   potential newly discovered moving objects are characterized by very few
   observations, acquired over a limited time.
	 Neither can those objects be specifically targeted
   by \gaia itself
   after their first detection.
   This aspect was recognized early in the design of the \gaia data processing. 
   }
   {A daily
   processing pipeline dedicated to these candidate discoveries
   was set up to release calls for
   observations to a network
   of ground-based telescopes. Their aim is to acquire follow-up astrometry and
   to characterize these objects.
   }
   {From the astrometry measured by \gaia, preliminary orbital solutions are 
   determined, allowing to predict the position of these potentially
   new discovered objects in the sky
   accounting for the large parallax between \gaia and the Earth (separated by
   0.01\,au).
   A specific task within the \gaia Data Processing 
   and Analysis Consortium (DPAC) has been responsible for the distribution of 
   requests for follow-up observations of potential \gaia SSO discoveries.
   Since late 2016, these calls for observations
   (nicknamed \textsl{alerts}) are published via a Web interface with 
   a quasi-daily frequency, together with observing guides,
   freely available to anyone world-wide.
   }
   {Between November 2016 and the end of the first year of the extended mission
   (July 2020), over 1700 
   alerts have been published, leading to the successful recovery 
   of more than 200 objects. Among those, six have provisional designation
   assigned with the \gaia observations, the others being previously
   known objects with
   poorly characterized orbits, precluding identification at the time of 
   \gaia observations.
   There is a clear trend for objects with a high inclination to be 
   unidentified, 
   revealing a clear bias in the current census of SSOs
   against high inclination populations.
   }
   {}


\maketitle

\section{Introduction}

  The main science driver of the
	European Space Agency (ESA) \gaia astrometric mission
  is the study of the structure and the 
  dynamics of the Milky Way \citep{2001AA...369..339P}.
  Building upon the heritage of HIPPARCOS, 
  \gaia was designed to conduct a survey of the
  full celestial sphere \citep{2016AA...595A...2G}, at an absolute 
  precision of 25 micro-arcseconds ($\mu$as) for the parallax for V=15 mag,
  solar-type stars, and 13 $\mu$as/yr for the proper motion.
  \gaia's on-board image processing detects and measures all celestial sources 
  brighter than $\approx$20.7 mag in G, 
  \gaia wide visible filter \citep{2010AA...523A..48J}.
  It also measures the photometry in the G filter 
  and
  low-resolution spectroscopy for all sources.  

  To achieve such accuracies, \gaia eliminates systematic errors by
  adopting the same strategy as its predecessor HIPPARCOS,
  i.e., by measuring simultaneously with high precision the epochs of transit of all sources in two
  fields of view separated by a wide angle
  (the {\sl Basic Angle}) of 106.5\degr, during a continuous scan at a
  constant rate (period of 6\,h). 
  The practical realization (e.g., pixel scale, binning)
	implies a widely different astrometric precision between the
  ``\textsl{along scan}'' (AL) direction
  (tangent to the great circle scanned by the rotation) 
  and the perpendicular ``\textsl{across scan}'' (AC) direction. 
  
  Multiple transit observations of the same portion of the sky, with varying
  scanning directions, are therefore required to measure the positions 
  at the required $\mu$as precision. 
  This is realized through the precession of the spinning
  axis\footnote{which is anyway required to cover the entire celestial sphere.}
  in 64 days, always pointing 45\degr~away from the Sun.
    The so-called Astrometric Global Iterative Solution (AGIS) produces 
  the astrometric model of the whole sky, corresponding to the combination 
  of all star positions and proper motions, plus calibration parameters. 
  For a complete description of the satellite and its operation 
  we refer the 
  interested reader to \citet{Prusti16}; below we
  summarize some features that are relevant for the present work. 

  The two 1-m telescopes of \gaia are mounted on a structure pointing 
  to the two fields separated by the Basic Angle, and produce an image on 
  a single focal plane. The CCDs composing the focal plane operate in 
  Time-Delayed Integration mode (TDI), in which electrons are
  moved from pixel to pixel at the same rate as the sources (hence the
  photo-electrons) drift along the CCD pixel lines
  (i.e., the AL direction).

  \gaia's focal plane contains several instruments, each corresponding
  to different ``\textsl{strips}'' of \numb{7} CCDs each.
  First, the SkyMapper (SM)
  instrument identifies the sources from each telescope and discriminates
  their origin between the two fields of view, by two CCD strips.
  It is followed by nine CCD strips for astrometry
  (the Astrometric Fields, AFs). Hereafter, \textsl{observations}
  refer to the positions on each of these 9 CCDs, while \textsl{transit}
  encompasses these observations,
  as in \citet{Spoto18}.
  Then two strips provide slitless low-resolution
  spectroscopy, splitting the visible spectrum in a blue and a red component,
  named BP and RP \citep{2018AA...616A...3R}.
  A more restricted portion of the focal plane is devoted to the three CCD 
  strips of the radial-velocity spectrometer RVS \citep{2018AA...616A...5C},
  collecting high-resolution spectra of bright stars (V<17).
  With each of these 16 CCD strips containing 7 CCDs 
  (except the 3 RVS strips with only 4 CCDs each) and a few technical CCDs,
  \gaia's focal plane is the largest ever operated in space
  in terms of the number of pixels.

  The resulting large data volume poses a challenge for the telemetry. 
  Since the start of its operations in July 2014, \gaia has been located
  at the Sun-Earth L2 Lagrangian point (0.01\,au from the Earth). Upon detection by the SM,
  only small windows around each source 
  are tracked along the focal plane.
  Most frequently, these windows have 6$\times$12
  pixels\footnote{The angular size on the sky of the pixels of the Astrometric
  Fields is 59$\times$177 mas.}
  only.
  Most of them are also binned in the AC direction so that only a 
  unidimensional signal is transmitted to the ground. 
  A notable exception is for sources brighter than G$<$13,
  for which larger 2-D windows are preserved. 

  All the aforementioned characteristics of \gaia have strong consequences
  on its observation of Solar System Object (SSO) transits such as:
  \begin{itemize}
    \item[$\bullet$] Source tracking on the focal plane follows closely
      the rate of stars. For this reason, SSOs may drift with respect 
      to the centre of the assigned windows in the AF, and ultimately 
      leave it if the apparent velocity is large enough.
      A typical main-belt asteroid drifts by 1 pixel in AL during a 
      transit over a single CCD (4.4\,s) if it moves at a typically
      rate of 13~mas/s
      along the same direction. 
    \item[$\bullet$] The motion of the SSOs produces a distortion of
        the signal, the shape of which is no longer well-represented
        by a pure stellar point-spread function.
    \item[$\bullet$] As a consequence, the astrometric and photometric
      processing must be adapted to cope with the resulting flux loss,
      variable and increasing over the transit.
    \item[$\bullet$] Each measured position of an SSO is strongly constrained 
      in the AL direction (at the 0.1--20 mas level, depending on the astrometric 
      solution used to reduce the data, see below) but much worse in the AC
      direction ($\sim 600$ mas). 
    \item[$\bullet$] The identification of SSOs, due to their motion relative 
      to stars, cannot be determined by the internal cross-matching of sources 
      \citep[which is the root of the creation of the stellar
        catalog,][]{2016AA...595A...3F} but must rely on a specific processing
      requiring an external catalogue of orbits.
  \end{itemize}
  
  The potential for Solar System research of the absolute and extreme-precision 
  astrometry, photometry, and low-resolution spectroscopy of \gaia was 
  recognized early on \citep{2007EMP..101...97M}.
  Furthermore, early estimates considering the limiting magnitude of 
  \gaia and the completeness
  of SSO catalogs, predicted about a hundred discoveries of SSOs per week 
  \citep[mainly main-belt asteroids, with potentially a couple of near-Earth 
  asteroids, and no objects in the outer Solar
  system:][]{2007EMP..101...97M, 2014-GFS-Carry}.
  
  While large ground-based surveys (e.g., Pan-STARRS, Catalina Sky Survey, 
	Legacy Survey of Space and Time)
  are expected to discover and gather multiple observations of most objects,
  the specific location of \gaia and the small solar elongation (45\degr) reached by 
  its observations allows the discovery of near-Earth asteroids (NEAs) with 
  small aphelion distances.

  Within the \gaia Data
  Processing and Analysis Consortium (DPAC),
  the potential for SSO discoveries led to a
  specific workflow for SSOs, consisting in two chains. 
  A short-term processing pipeline (named SSO-ST) runs daily to identify SSOs in the
  latest (72\,h) transits \citep{Tanga16}, and to trigger follow-up
  observations from the Earth.
  A ground-based follow-up network of telescopes
  (\gaia-FUN-SSO) was established to ensure 
  observations and confirm \gaia discoveries \citep{Thuillot18}.
  The long-term processing pipeline, named SSO-LT, 
  benefiting from the full global astrometric solution of \gaia,
  runs on longer timescales to produce the catalogues
  with the best astrometry, photometry, and spectroscopy of SSOs for the various
  Data Releases \citep[DRs, of which the first with SSOs was \gaia DR2,
  see][]{Spoto18}.
 
  We focus here on the results of the SSO-ST over the course of the \gaia nominal
  mission (July 2014 -- July 2019)
  and the first year of the extended mission (July 2019 -- July 2020).
  The article is organised as follows.
  \Autoref{sec:ssost} summarizes the main steps of the daily processing 
  \citep[beyond the description of the
   chain provided by][]{Tanga16}.
  \Autoref{sec:du459} presents the system through which calls for
  observations of potential discoveries are released.
  The network of ground-based telescopes is presented in \Autoref{sec:gfs},
  and a summary of the successful SSO recoveries in
  \Autoref{sec:dist}.
  We discuss the orbital properties of the alerts in 
  \Autoref{sec:orbit}.

\section{The SSO-ST daily processing\label{sec:ssost}}

  We briefly summarize the daily processing of \gaia observations in the
  framework of the SSO-ST. For a full description of the pipeline, we refer to
  \citet{Tanga16} and \gaia
  documentation\footnote{\url{https://gea.esac.esa.int/archive/documentation/}}.
  We focus here on its differences with the long-term processing,
  which is thoroughly documented in \cite{Spoto18}.
 
\begin{figure}[t]
\centering
  \includegraphics[width=0.50\textwidth,clip]{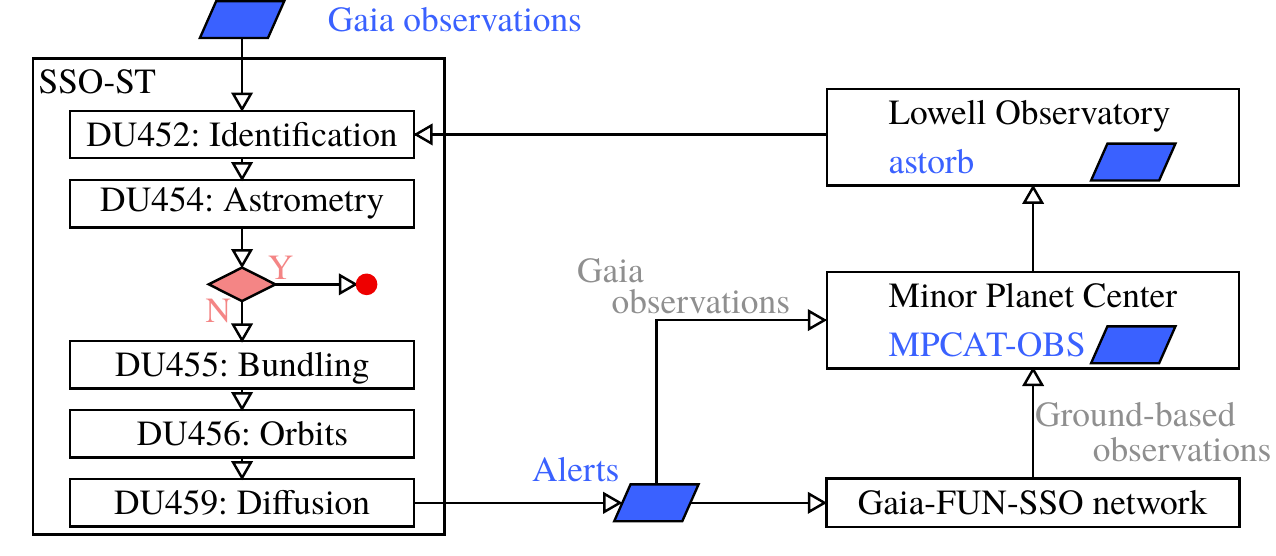}    
  \caption{%
    Simplified workflow from \gaia observations to alert dissemination, 
    reporting to the Minor Planet Center, and update of \astorb, 
    the database of orbital elements used for the identification.
  }
\label{fig:astorb}
\end{figure}

  An Initial Data Treatment (IDT) of the \gaia data is performed upon reception on 
  Earth \citep{2016AA...595A...3F}. After obtaining the signal parameters that 
  describe the position of a source on the focal plane at the observing epoch,
  IDT uses a preliminary 
  great-circle attitude solution to derive its position on the sky,
  and executes a first cross-matching to sources previously observed. 
  Typical sources that fail cross-matching are either artifacts 
  (e.g., cosmic ray, diffraction spikes from bright sources, spurious detections), 
  sources at the limit of detection (detected only on certain transits), 
  or transients such as genuine moving objects from our Solar System.
  The position of the latter always changes on the sky.
  In these cases, the unmatched sources are removed from the main stellar data 
  processing, and are available for a special processing for SSOs
  (\Autoref{fig:astorb}).

  The first task of this processing is to identify if the source corresponds to
  a known SSO (task performed by the Development Unit DU452 of \gaia DPAC).
  Potentially, all unmatched sources are SSO candidates. However, SSOs constitute
  a very minor fraction of all the sources continuously transiting 
  the focal plane of \gaia.
  By taking a somewhat optimistic number of 350,000 asteroids that \gaia can
  observe during the entire duration of its mission, \gaia observes on average
  1 SSO for every 4,000 stars.
  It is clear that any inefficiency in the IDT cross-matching for stars
  can produce an overwhelmingly large number of false SSO detections
  (for instance, at a 99.9\% efficiency, there are 4 unmatched stars for each 
  genuine SSO).    
 
  Furthermore, a very large number of contaminants at the CCD level
  was found to heavily populate
  the sample of unmatched detections, making the task of SSO 
  identification impossible without 
  appropriate cleaning of the data set. The appropriate filtering was made 
  possible by introducing in IDT the computation of the AL velocity of the 
  source (and AC whenever possible) on the focal plane.
  All sources without a detectable motion are discarded. 
  The threshold is dynamically adjusted on the base of the velocity distribution 
  for each 1-day data chunk and is typically of the order of $\pm$2\,mas/s.
 
  For an object that passes through the filter, its positions are checked
  against the predicted positions of known SSOs at the corresponding epochs. 
  \numb{Weekly}, the catalogue of osculating orbital elements of the minor planets
  \astorb\footnote{Based on the world-wide  
    catalog of observations maintained at the 
    Minor Planet Center
    \citep[MPCAT-OBS,][]{Rudenko16}.}
  \citep{Bowell14,Bailen20} is updated,
  the ephemerides for all SSOs are pre-computed and
  stored in a database \citep[this is a version of the
  SkyBoT software dedicated to \gaia, see][for more details]{
    2006-ASPC-351-Berthier, 2016-MNRAS-458-Berthier}.
  If not linked with a known SSO, the source may still be either a known SSO 
  with a poorly characterized orbit (which precludes identification), 
  an unknown SSO, or an artifact as listed above.
  Because the identification relies on an external catalogue, it is crucial to 
  keep the latter up to date. Similarly, any bias in the current census of SSO populations 
  will affect the identification.

  A suite of different tasks are then performed in chain on each unidentified source.
  The suite of observations (i.e., SM+AF, up to ten) defining a transit is 
  converted into sky coordinates by DU454, after being subjected to a quality filtering
  to mitigate the distortion effects, due to the motion, on the
  accuracy of the centroid determination (DU453).
  The only attitude (the transformation from pixel coordinates to sky coordinates) 
  available hours after the observations is the \textsl{poor} 
  (by \gaia's standards) Oga1 attitude,
  containing many irregularities. 
  One of the tasks of DU454 is to smooth the attitude and 
  remove these irregularities, which is necessary to check whether a 
  linear motion in the sky can be fit to the transit.  
  The obtained smoothed attitude has typical uncertainties of 25 mas 
  in AL and 40 mas in AC, which is still much larger than the uncertainties 
  on the high-quality Oga3 attitude that will be determined 
  by AGIS, months later (hence too late to be used for alerts) 
  for the data releases. 
  Next, DU454 will try to fit a linear motion on all positions of 
  a transit, and remove those that do not fit on the linear motion, 
  and that are either of too bad quality, or not detections of the SSO.
  If no linear motion can be fit at all on the transit, the entire transit 
  is rejected, and is supposed to be a contaminant.
  
\begin{figure}[t]
\centering
  \includegraphics[width=0.50\textwidth,clip]{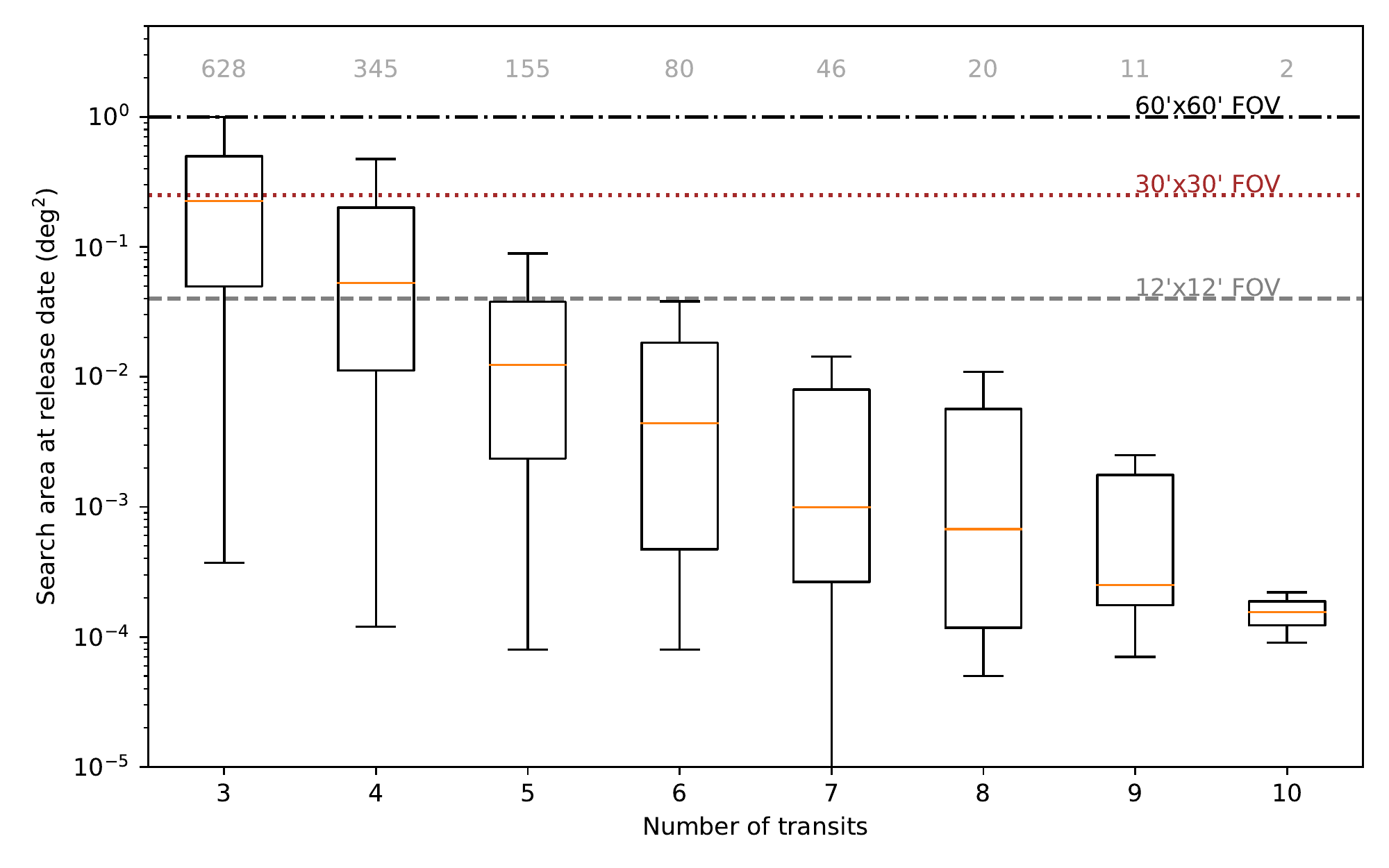}
  \caption{Area on sky (present as 25-50-75\% whiskers and 3\,$\sigma$ min,max)
	  covered by the preliminary orbits as function
    of the number of \gaia transits used. The gray numbers a the top correspond
    to the number of alerts released with the associated number of transits.}
\label{fig:area}
\end{figure}

  A procedure (DU455) then attempts to link together transits, in order\
	to identify which ones
  belong to the same source.
  Based on current census of asteroids, their typical
  apparent motion on the sky as a function of solar elongation
  (the most important parameter for this task) is determined.
  The linking procedure then performs an efficient search to find
  all possible links that satisfy this typical motion.
  The result is further filtered on similarity of the measured
  apparent magnitude, and compatibility of the apparent motion with 
  the instantaneous velocity produced by IDT.
  This process is applied daily on the last 48 hours worth of data.
  Once at least two transits are linked together, the source may be
  a genuine moving object. However, a few transits by \gaia
  \citep[two or three only, in 45\% and 10\% of the cases, providing
	less than 6\,h of observations,][]{2011sssb.confE...2T}
  are not enough to compute an accurate orbital solution.
  The uncertainty associated to these preliminary orbit solutions are inversely 
  proportional to the 
  number of transits (\autoref{fig:area}).

  A short-arc solution, valid for a limited interval of time, is, however,
  required for ground-based follow-up, which is essential to secure the orbit.
  The proximity of SSOs compared to the Earth-\gaia distance
  implies large parallaxes: from an arcminute for the distant Kuiper
  belt at 30\,au from \gaia to over a \textit{degree} for nearby
  (less than 0.5\,au) Mars-crosser and near-Earth asteroids.
  Because the distance of SSOs is unknown upon first observation, the parallax
  cannot be accounted for to convert the coordinates measured by \gaia
  into Earth-based coordinates.
  Preliminary orbits based on \gaia short arcs are thus computed
  within the SSO-ST (in the processing unit named DU456) to allow short-term predictions 
  \citep{2016PSS..123...95M} and to guide ground-based observations.
  These preliminary orbits are determined through a 
  random-walk independent sampling of ranges of SSOs.
  This provides \mcmc candidates orbits
	(represented by Keplerian elements) for each new source. 
  We refer to \cite{2018-AA-620-Fedorets} for details.
  
  These short-term orbital elements, improved by 
  ground-based follow-up, adds enormous value to \gaia observations.
  First, without follow-up observations, these \gaia observations 
  are ``lost'' as they cannot be linked with any known object.
  Second, without an orbit, the observed properties (position, apparent G magnitude,
  BP and RP spectra) of these asteroids cannot be appropriately studied 
  or understood in the context of SSO populations.
  Such situations can arise in modern surveys and can lead to a significant 
	loss of valuable data. 
  As a matter of example, the latest release
  (2008) of the moving object catalog
  of the Sloan Digital Sky Survey \citep{2001AJ....122.2749I}
  contains over \numb{470,000} observations of moving objects of which
  only about \numb{220,000} are linked with known SSOs.

  The SSO-ST was set up to avoid such a situation. 
  Its main goal is to release calls for observations
	(\textsl{alerts}) 
  of these unidentified sources, allowing for observations from the Earth 
  and hence the determination of a preliminary orbit, and update of the 
  database of orbits (\autoref{fig:astorb}).

  The pre-launch plans for the SSO-ST were to process all SSOs 
  with at least two transits by \gaia.
  The majority of cases represent the situation where
  only two transits exist for an object.
  However, the spread of orbital solutions for these cases
  typically results in a large spread of solutions on the sky.
  This renders any follow-up efforts unfeasible.
  Therefore, we operate the pipeline for objects with a
  minimum of \numb{three} transits (\Autoref{fig:area}).

\section{The alert release interface\label{sec:du459}}

\begin{figure}[t]
\centering
  \includegraphics[width=0.50\textwidth,clip]{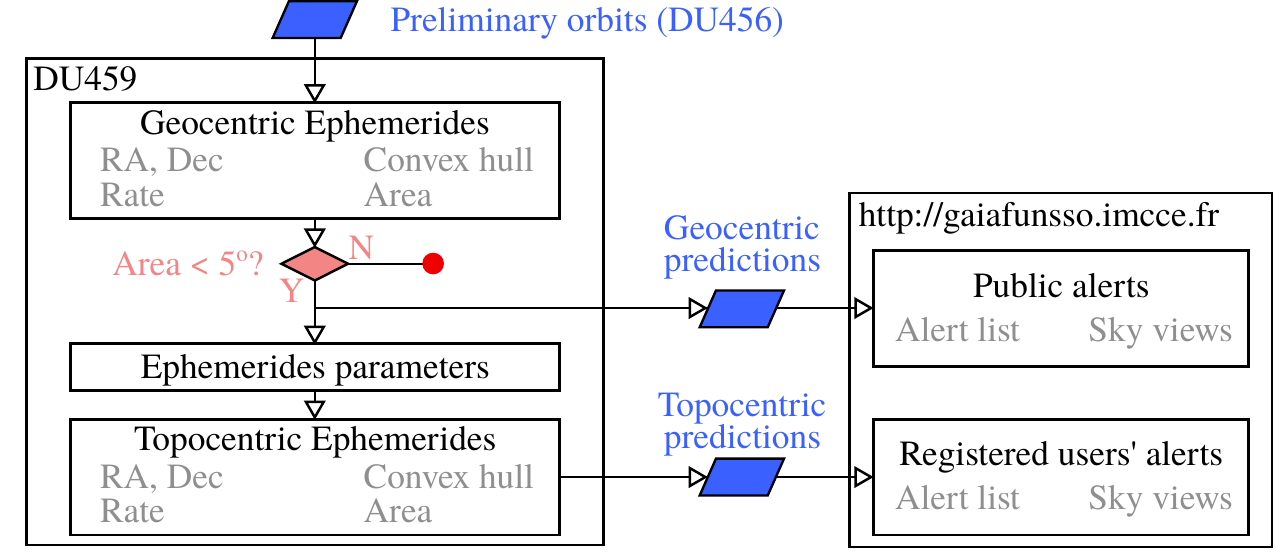}
  \caption{Simplified workflow of alert processing and diffusion.
    Alerts are selected based on geocentric ephemerides. For each alert
    below the area threshold, topocentric ephemerides are computed for 
    each observatory in the system. All ephemerides are stored in a database
    which is used by the Web pages.}
\label{fig:du459}
\end{figure}

  An interface to release the calls for observations (\textsl{alerts})
  to the community of observers was set up.
  This is the last task of the SSO-ST (DU459), the workflow of which
  is described in \autoref{fig:du459}. 
  
  All sources are treated independently.
  For each, the geocentric ephemerides of the \mcmc orbits are computed for  
  30 days starting at the current epoch, with a time step of one day.
  At each computed epoch, the different orbits result in a cloud of 
  different positions on the sky.
  The median position with the estimated apparent velocity vector, 
  the convex hull of the cloud, and its angular surface
  are computed at each time step,  
  and stored in a database.
  
  Given the spread in the initial orbital elements, the dispersion of the cloud of
  predicted positions increases over time. When the search area has grown 
  beyond a certain size, a very large field of view or a time-consuming
  search strategy would be required. For this reason, 
  the alerts are considered to be valid only between the epoch of computation and the
  epoch at which the area on sky becomes larger than \numb{five} square
  degrees.
  If at the first epoch, the area is already larger than this threshold,
  the corresponding alert is skipped and the pipeline processes the
  following entry in the alert list. 

  Topocentric ephemerides (for each observatory registered in the system, 
  see below) are then computed.
  The time step and duration of prediction are adapted to each observatory,
  according to each user's preference (i.e., maximum search area on the sky).
  Similarly to the geocentric ephemerides, the area, median position, 
  and convex hull of the cloud of predictions are computed for each alert, 
  observatory, and time step; 
  and stored in a database.

\begin{figure}[t]
\centering
  \includegraphics[width=0.45\textwidth,clip]{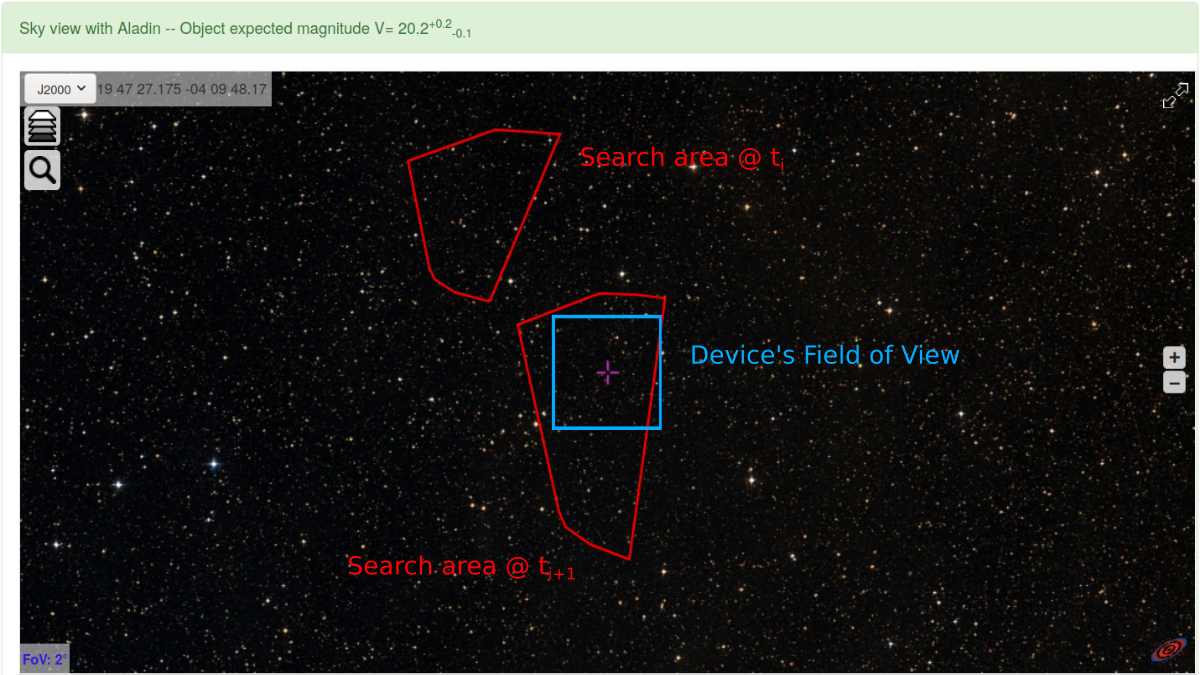}  
  \caption{Example of the sky view of alerts with \aladinlite.
  The red polygons correspond to the convex hull of the cloud of predicted position
  at different dates, and the blue square represents the field of view of the
  device defined by the user (only available upon registration).}
  \label{fig:gfs:sky}
\end{figure}

  All these predictions are published online through a suite of online
	portals\footnote{\url{https://gaiafunsso.imcce.fr/}}.
  A public page lists all the alerts
  whose area is smaller than \numb{one} deg$^2$
  and the apparent magnitude brighter than \numb{V=21},
  based on their geocentric ephemerides.
  For each alert, the dates of release and end of validity are given, together with
  an identifier\footnote{The nomenclature is \texttt{gYwNNN}, with
  \texttt{Y} the year of \gaia operations,
  \texttt{w} the week of the year, 
  \texttt{NNN} the incremental number of alerts in that week.}, 
  the predicted median Right Ascension and Declination,
  area,
  and apparent magnitude. 
  The details of each alert are accessible on specific pages, in 
  particular a display of the convex hull of the predictions at each
  time step with \aladinlite
  \citep[\autoref{fig:gfs:sky},][]{2000AAS..143...33B}.

  These geocentric predictions are available to everyone.
	More functionalities are, however, provided to registered users.
  Anyone can register for free to the system,
  by filling a simple form.
  Upon registration, users can easily list their observing devices
  (i.e., observatories).
  The main characteristics of each device are its location on Earth
  (either by entering its longitude, latitude, altitude, or 
  IAU MPC observing
  code\footnote{\url{http://vo.imcce.fr/webservices/data/displayIAUObsCodes.php}}),
  its field of view, and the thresholds in
  apparent magnitude and area on sky to consider for alerts. 
  Hence, for registered users, the list of alerts proposed by the Web 
  pages contains only those fulfilling the user's observability criteria:
  apparent magnitude brighter than the threshold, 
  area on sky smaller than the threshold,
  declination observable from the observatory.
  Furthermore, the pages presenting the details provide additional features:
  \begin{itemize}
    \item[$\bullet$] the \aladinlite sky view shows the field of view of the device, to help
      preparing the observations (\autoref{fig:gfs:sky});
    \item[$\bullet$] the convex hull at each time step can be downloaded (to be used in
      \aladin or user in-house software); 
    \item[$\bullet$] a simple page is proposed to observers to report on the
      status of their observation (success or target not found).
  \end{itemize}

\section{The \gaia-FUN-SSO ground-based network\label{sec:gfs}}

\begin{figure}[t]
\centering
  \includegraphics[width=0.5\textwidth,clip]{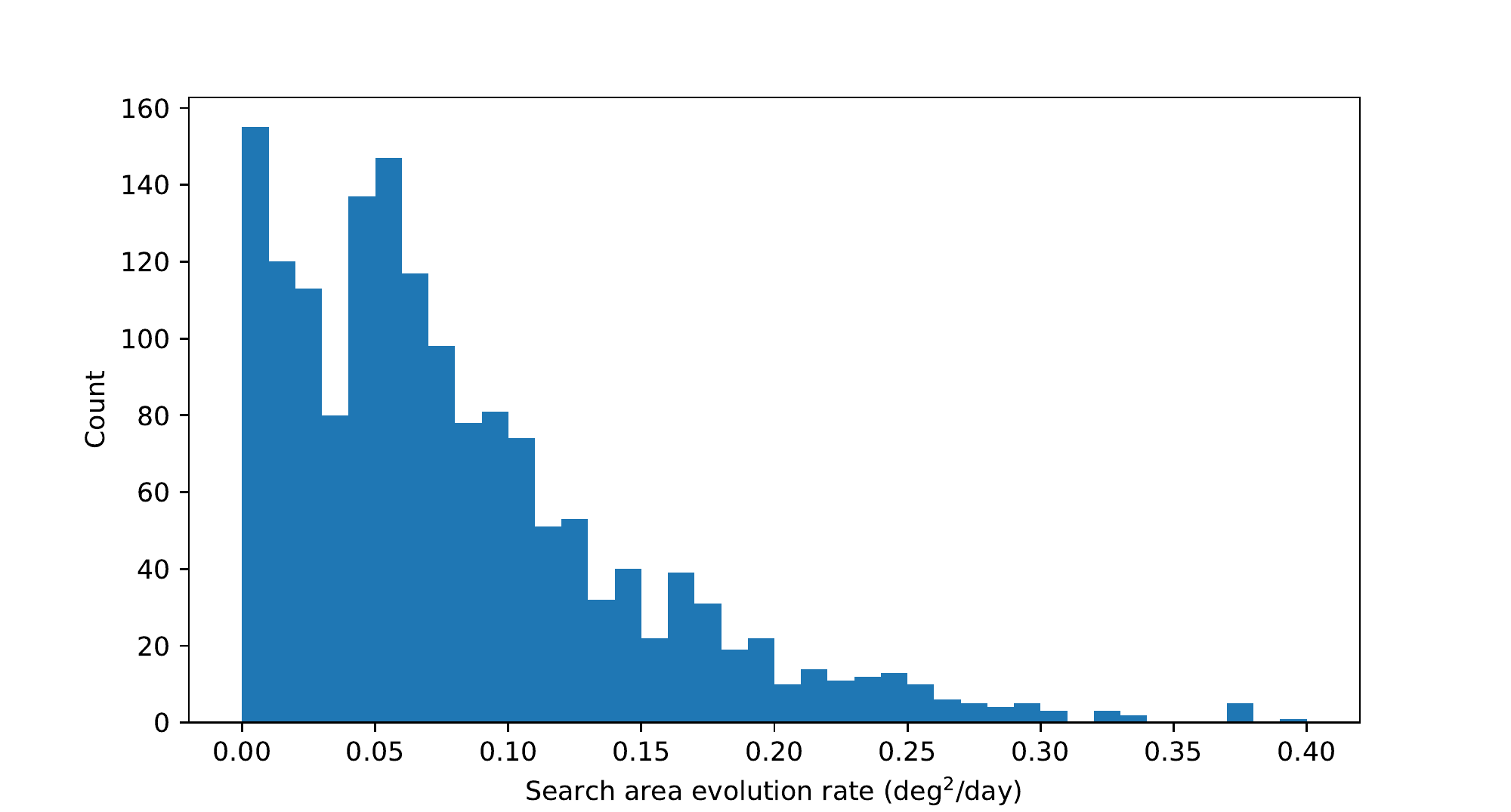}  
  \caption{Distribution of the growth rate of the search regions, in
    deg$^2$/day, for all alerts released since November 2016.
    The median growth rate is 0.06\,deg$^2$/day with a standard
    deviation of 0.08\,deg$^2$/day.
    }
  \label{fig:histrate}
\end{figure}

  As explained in \Autoref{sec:ssost}, the necessity of follow-up observations from
  ground-based stations was recognized early on.
  Two aspects, closely related to the specificity of \gaia, required multiple
  stations, widely spread over the globe to cover a large range of longitude
  and latitude.
  First, \gaia is an all-sky surveyor, hence requiring follow-up stations in both
  the northern and southern hemisphere. 
  Second, the strong yet unknown parallax between \gaia and the Earth carries a
  significant uncertainty on the sky coordinates of any new detections
  \citep{2012PSS...73...21B}, which
  increases with time (\Autoref{fig:histrate}).
  Observations as early as possible after detection by Gaia
  are therefore more likely
  to succeed than delayed ones, calling for a wide coverage in longitudes.

  Before the launch of \gaia, efforts were conducted to build this large
  network. The early assessment of the magnitude range of potential
  discoveries opened the possibility
  of alerts observable with modest apertures (0.5 to 1\,m).
  Our requests for volunteers were warmly received, and over \numb{150}
  participants, including amateur astronomers,
  had registered in the alert release system by the time of
  \gaia's launch.
  We consolidated and interacted with the network of volunteers (observations
  were to be conducted on the best-effort basis) through three workshops,
  held in Paris Observatory in 2010, 2012, and 2014
  \citep{2010-GFS-WS,2012-GFS-WS,2014-GFS-WS}.
  We also trained the network to react to alerts by releasing calls for 
  observations on NEAs \citep{Thuillot15b}.
  
\begin{figure}[t]
\centering
  \includegraphics[width=0.5\textwidth,clip]{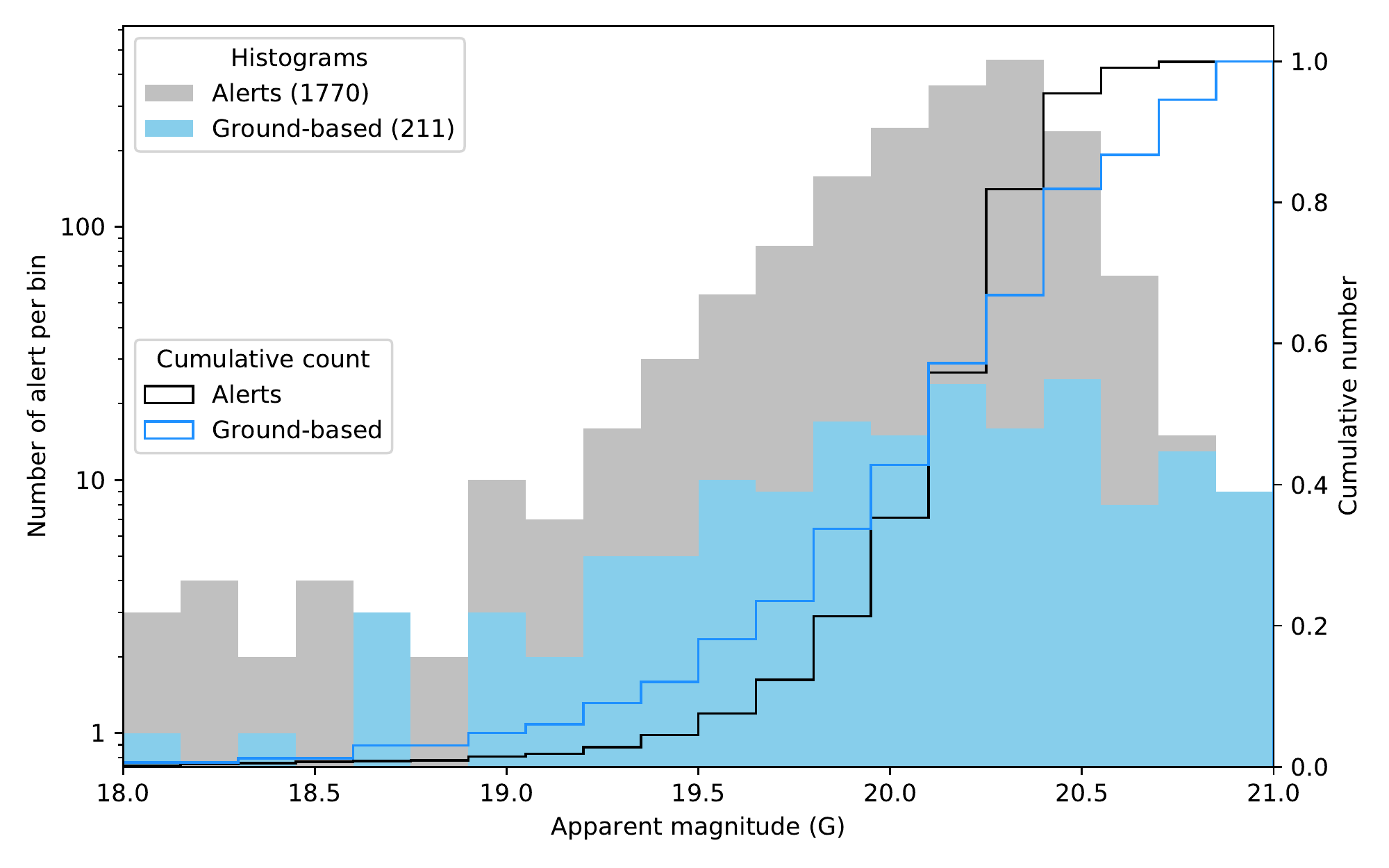}    
  \caption{Distribution of the G magnitude of all alerts 
    and of ground-based observations.}
\label{fig:mag}
\end{figure}

\begin{figure}[t]
\centering
  \includegraphics[width=0.5\textwidth,clip]{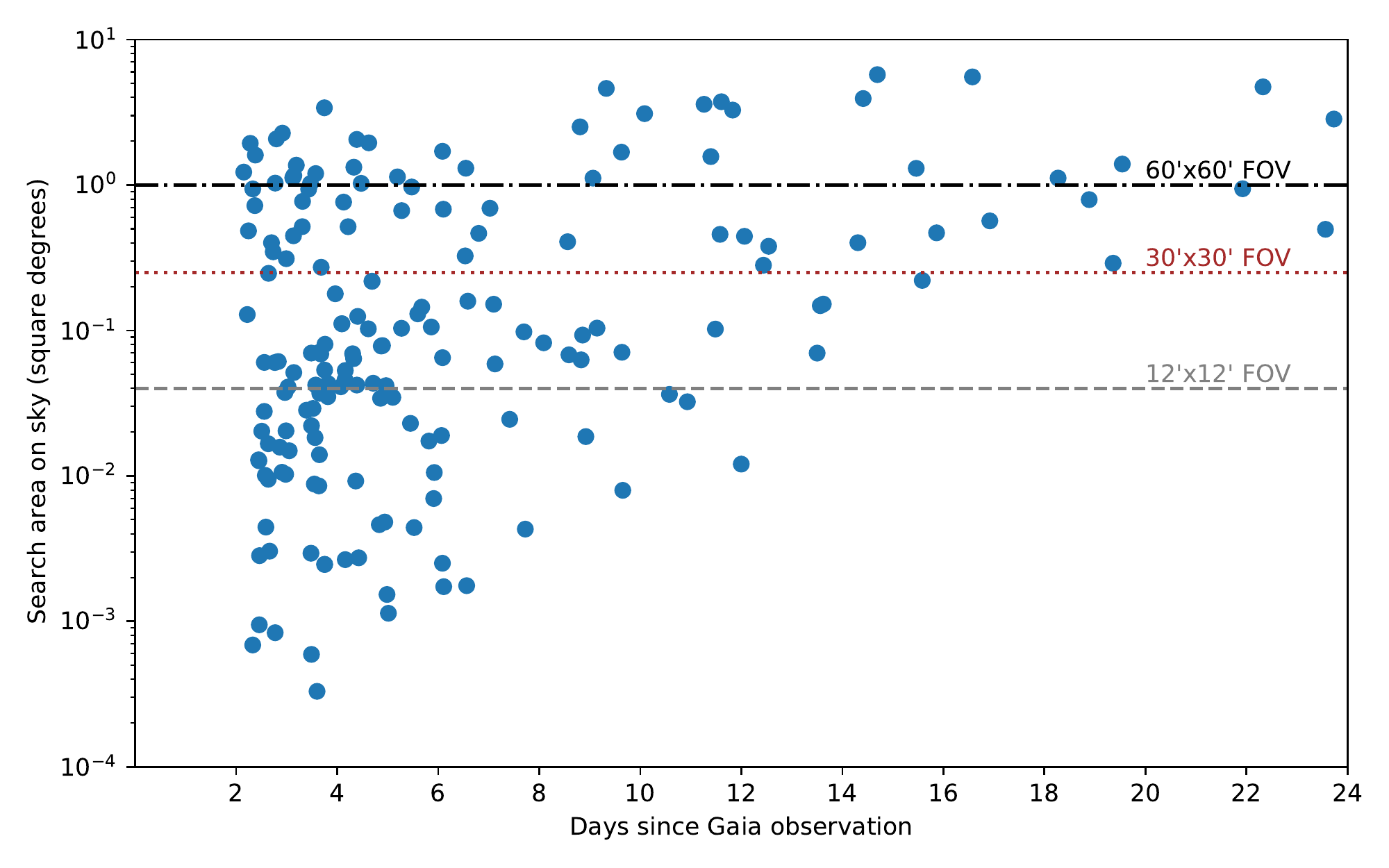}    
  \caption{Extent (in square degrees) of the search regions
	  of alerts at the time of their observations
    by the network, counted in days since \gaia observations.
    }
\label{fig:areatime}
\end{figure}

  However, the contribution of the network has been concentrated in only a few
  observatories since 2016, with fewer detections per week than foreseen.
  First, between the early assessment of the \gaia capabilities
  for SSOs in 2007, the start of the \gaia operations in 2014, and the effective
  daily processing of alerts in November 2016 (see below), 
  wide surveys (Pan-STARRS, WISE) had discovered most of the
  SSOs observable by \gaia (V\,$\leq$\,20.7).
  The bulk of alerts hence corresponds to objects fainter than 
  V\,$\approx$\,20, seldom brighter (\autoref{fig:mag}).
  Telescopes larger than typically 1\,m are therefore required, and
  cameras with large field of views are favoured, which 
  strongly limits the potential contribution by amateurs.
  Second, the areas to search within increase with time due to the 
  lack of constraints on the short-term orbit
  (\autoref{fig:histrate}, \autoref{fig:areatime}).
  Owing to delays from the observation on-board to the downlink to Earth,
  to the initial data treatment and finally to the many steps in the 
  SSO-ST, alerts are released at the earliest about \numb{48}\,h after
  the observations. The area and its evolution with time are strongly
  tied to the number of transits observed by \gaia.
  With increasing delay, it is often required for observers to scan the
  area to search with multiple exposures as it becomes larger than their
  instrument field of view.
  
  Although that fact limited the kind of facilities participating to the network, the
  numerous discoveries by other surveys did not reduce the interest in the 
  SSO-ST.
  First, \gaia, being an all-sky survey observing
  at solar elongation between 45\degr and 135\degr, provides the opportunity
  to discover SSOs in region of the sky poorly covered by ground-based telescopes.
  Second, SSOs with poorly characterized orbits are likely not recognized and 
  hence are processed by the SSO-ST as new sources. Observations by the
  network provide new constraints which improve their orbit, allowing 
  the MPC to link them with known SSOs, and hence a 
  subsequent identification
  by \gaia or by ground-based observatories.


\section{Results\label{sec:dist}}

  Soon after the start of \gaia's regular operations in September 2014
  \footnote{The scanning law during the first three months,
    called Ecliptic Pole Scanning Law (EPSL), was different
    \citep{2016AA...595A.133C}},
  the SSO-ST faced several issues, which did not show up in the pre-launch
  simulations. The major issue which delayed the release of alerts was
  the number of contaminants, several orders of magnitude above expectations.
  Filtering these contaminants required several cycles of tests and
  adaptations of the SSO-ST pipeline, so that the pipeline became fully functional no earlier than
  November 2016, which is a delay of 24 months with respect to the expectations. 
  Since then, the pipeline has been very robust, running continuously 
  on the computation facilities of CNES, in Toulouse, France,
  with the exception of a few technical breaks. It has also proved to produce
  a clean set of alerts in the output, effectively rejecting the contaminants.

  Since the start of the automated operations, more than \numb{1700} 
  alerts have been released. Observations from the ground have been performed for almost
  \numb{250} of them, leading to the successful observation of \numb{227}
  \gaia discovery candidates. Most of these observations were performed by a 
  small number of observatories (\Autoref{tab:gfs}) and an even smaller 
  number of astronomers (the observations at C2PU, OHP, and LCOGT being 
  carried out by the same observers, within a concerted effort to follow-up
  alerts from \gaia, for SSOs as described here but also photometric
  alerts, e.g., \citet{2019CoSka..49..420S}).

  This limited number of participants with respect to the large network built 
  before launch, is the result of two effects. First, the delay between 
  the launch of \gaia and the starting date of alert releases lowered 
  the interest of observers in the alerts.
  Second, the distribution of apparent magnitude and area of sky
  (\Autoref{fig:mag} and \Autoref{fig:areatime})
  limited the participants to those using large aperture telescopes
  with optical configurations providing large fields of view.

\begin{figure}[t]
\centering
  \includegraphics[width=0.50\textwidth,clip]{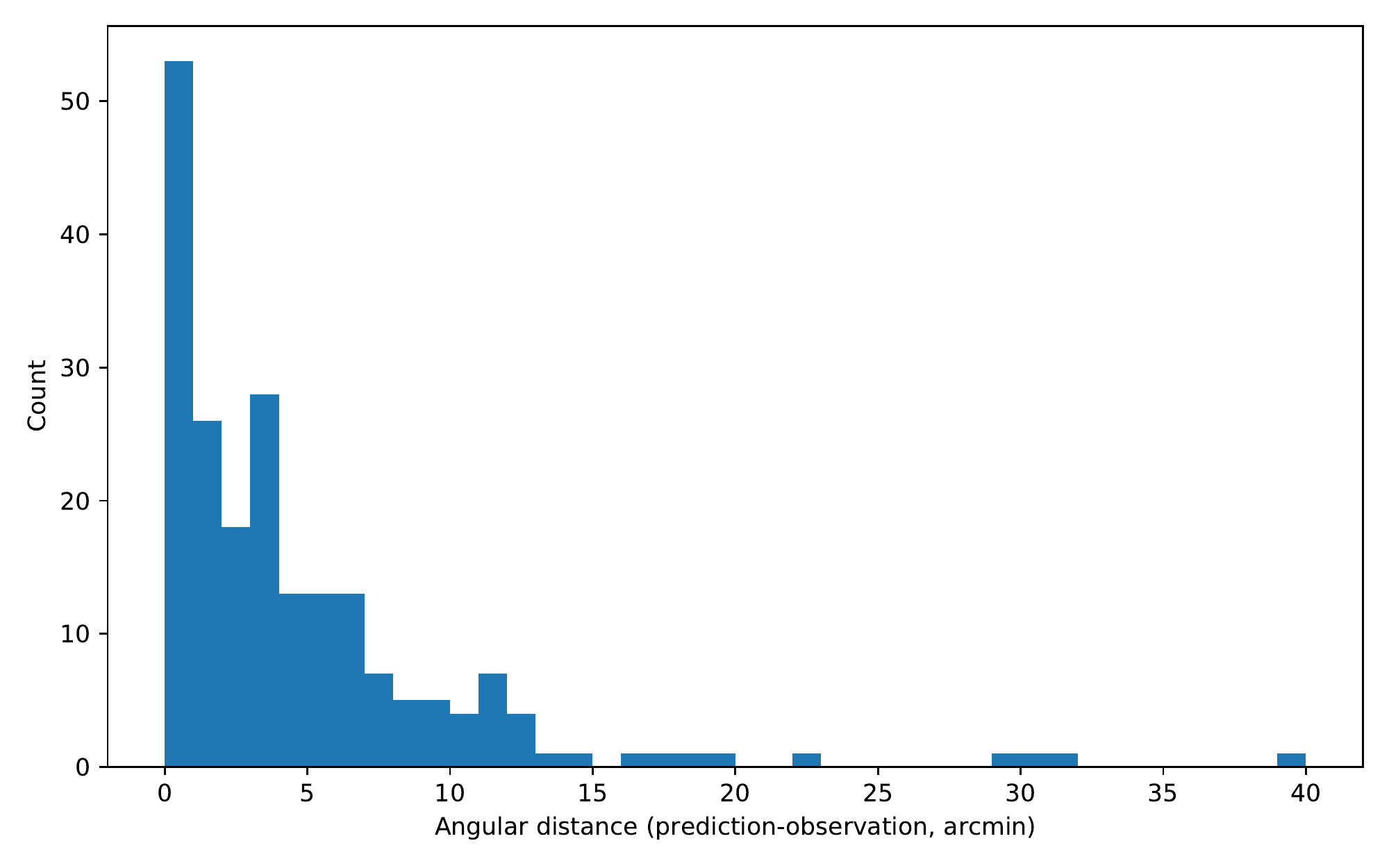}
  \caption{Distribution of the angular distance between the predicted
    median (RA,Dec) coordinates and the observed position measured
    from the ground.}
\label{fig:angdist}
\end{figure}

\begin{figure*}[ht]
\centering
  \includegraphics[width=1.\textwidth,clip]{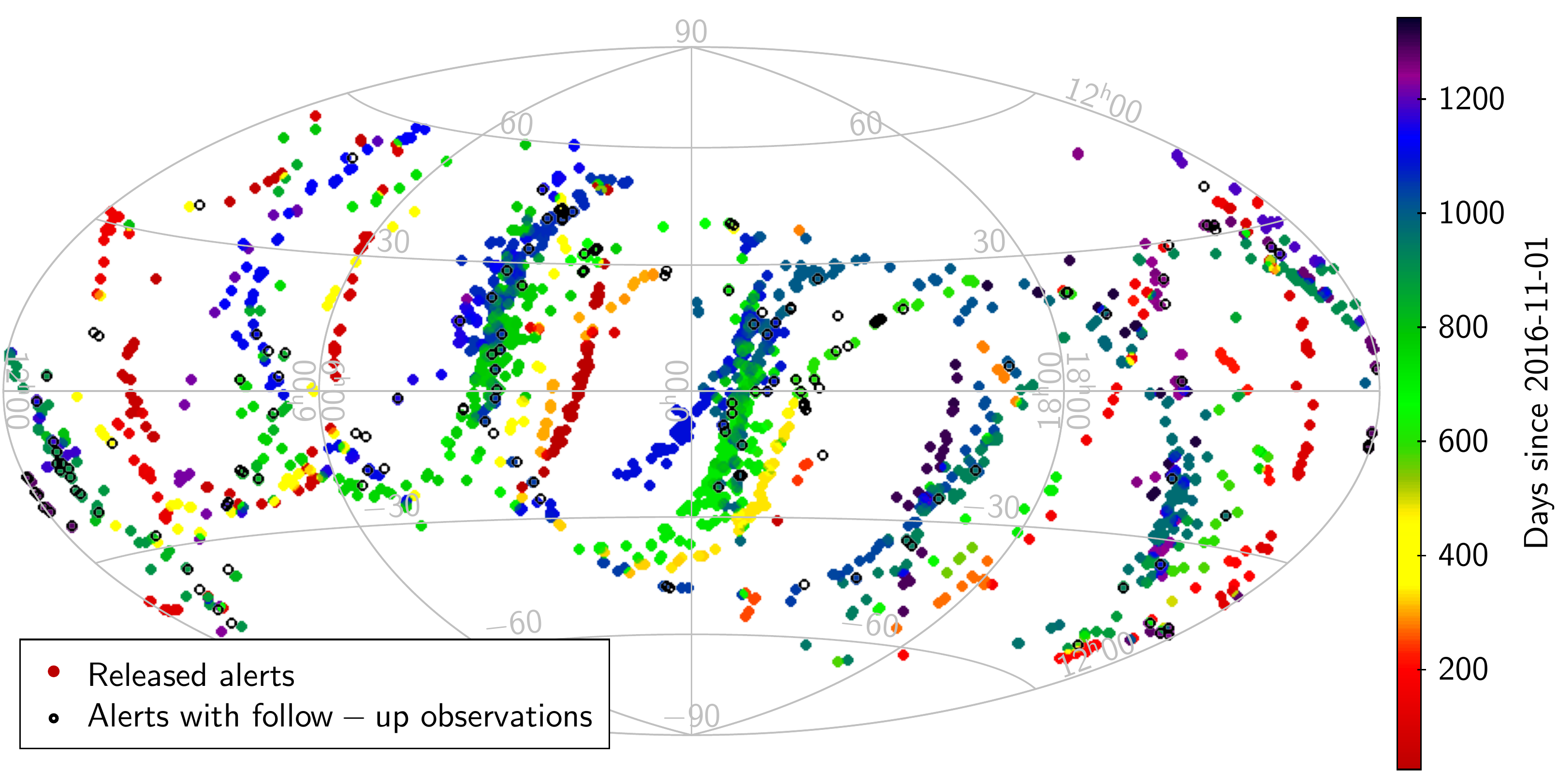}    
  \caption{Distribution of the alerts in equatorial coordinates
    (dots, colour-coded by epoch),
    together with the ground-based observations (open circles).
    The black line represents the ecliptic plane.}
\label{fig:eq}
\end{figure*}

\begin{table*}
\caption{Most-participating observatories of the network, with their
numbers of detection (\numb{up to July 2020}).}
\label{tab:gfs}
\centering
\begin{tabular}{lclcr}
\hline\hline
Observatory & Code & Country & Apert.(m)  & Detections \\
\hline
  Observatoire de Haute Provence (OHP)      &511 & France    &  1.2 & 44 \\
  C2PU, Observatoire de la C{\^o}te d'Azur) &010 & France    &  1.0 & 23 \\
  Terskol Observatory                       &B18 & Ukraine   &  0.6 \& 2.0 & 27 \\
  Kyiv Comet station                        &585 & Ukraine   &  0.7 & 21 \\
  Odessa Mayaki Observatory                 &583 & Ukraine   &  0.8 & 18 \\
  Abastumani Observatory, Tbilisi           &119 & Georgia   &  0.7 &  \\
  \multirow{10}{*}{
    \begin{tabular}{c}
        Las Cumbres \\
        Observatory\\
        Global \\
        Telescope\\
        Network
    \end{tabular}}  &V37 & USA         &  1.0 & \multirow{10}{*}{113} \\
                    &V39 & USA         &  1.0 \\
                    &W85 & Chile       &  1.0 \\
                    &W86 & Chile       &  1.0 \\
                    &W87 & Chile       &  1.0 \\
                    &Q63 & Australia   &  1.0 \\
                    &Q64 & Australia   &  1.0 \\
                    &K91 & South Africa&  1.0 \\
                    &K92 & South Africa&  1.0 \\
                    &K93 & South Africa&  1.0 \\
\hline
\end{tabular}
\end{table*}

  As such, most of the ground-based follow-up has been performed by the
  co-authors of the present article.
  Early observations were
  mainly performed at the OHP 1.2-m telescope, of which field of view
  (12$^\prime\times$12\arcmin) seldom covered the
  entire search region. Hence, we had to scan the search area with multiple
  exposures, slowing the process.
  We soon started to use the C2PU 1-m telescope, whose field of view
  (38$^\prime\times$38\arcmin)
  was more adapted to our needs.
  The successful recovery of the alerts
  g1P024
  and
  g1j03C (2012 VR$_{82}$),  
  at these telescopes in 2017 showed the pipeline 
  was delivering real SSO alerts which could be confirmed from the ground, starting 
  from the receipt of alert and up 
  to almost ten days after \citep{Carry19}.
  
  While the number of recoveries increased, we noted that 
  most objects are found well
  within the search area, on average at only 3\arcmin~from the median of the
  predicted coordinates (\Autoref{fig:angdist}). In other words, the
  area in the sky of alerts can be large, but 
  the reported medians (RA,Dec) are good estimators of the positions.
  Therefore, we continued to observe with the 1.2\,m at the OHP.
   Following these first recoveries, 
  the number of successfully observed alerts dramatically increased
  with the use of the Las Cumbres Observatory Global Telescope (LCOGT)
  network, which offers 1-m telescopes with large fields of view
  (27$^\prime\times$27\arcmin) spread worldwide, 
  hence perfectly adapted to the distribution of alerts on the celestial
  sphere (\Autoref{fig:eq}) and to the large search area.

  Before sending ground-based astrometry to the MPC, we always test whether
  the observation indeed corresponds to the object detected by \gaia. To this end, we
  compute an orbit based on both the \gaia transits and the ground-based astrometry
  \citep[the latter measured with the 
  \gaia Ground Based Optical Tracking software,][]{Bouquillon14}.
  We use a modified version of \texttt{OrbFit} \citep{orbfit}, used
  to validate the astrometry of SSOs in \gaia DR2 \citep{Spoto18}.

  The MPC has been collecting the minor planet astrometry acquired worldwide for decades.
  Upon reception of astrometry of a unidentified object, such as a \gaia
  alert, the MPC tries to link it with known objects, based on its
  observation database, including its unpublished part.
  Several cases are possible: the alert can correspond to a
  new object never observed before,
  an object recently detected by another observatory and not yet ingested in 
  the \astorb database, or a known object with a poorly determined orbit 
  (which precluded its identification within the SSO-ST and from the ground).

  The MPC is the sole international organization collecting astrometry,
  and we entirely rely on it to know if the \gaia alerts were hitherto unknown objects
  or recoveries of newly/poorly characterized SSOs.
  The \textsl{results}\footnote{\url{https://gaiafunsso.imcce.fr/stats/network.php}}
  page of the \gaia-FUN SSO Web 
  interface lists the designations of all recovered alerts.
  Of the \numb{250} follow-up observations, \numb{139} have received a designation, 
  but only \numb{6} are attributed to 
  \gaia (2018 XL$_{20}$, 2018 XL$_4$, 2018 YK$_4$, 2018 YM$_4$, 2019 CZ$_{10}$, 2019 HO$_4$).
  However, the others were poorly-characterized SSOs, and
  their observation in the context of the \gaia-FUN-SSO
  drastically improved their orbital elements.

\section{Orbital distribution\label{sec:orbit}}

\begin{figure*}[ht!]
\centering
  \includegraphics[width=0.95\textwidth,clip]{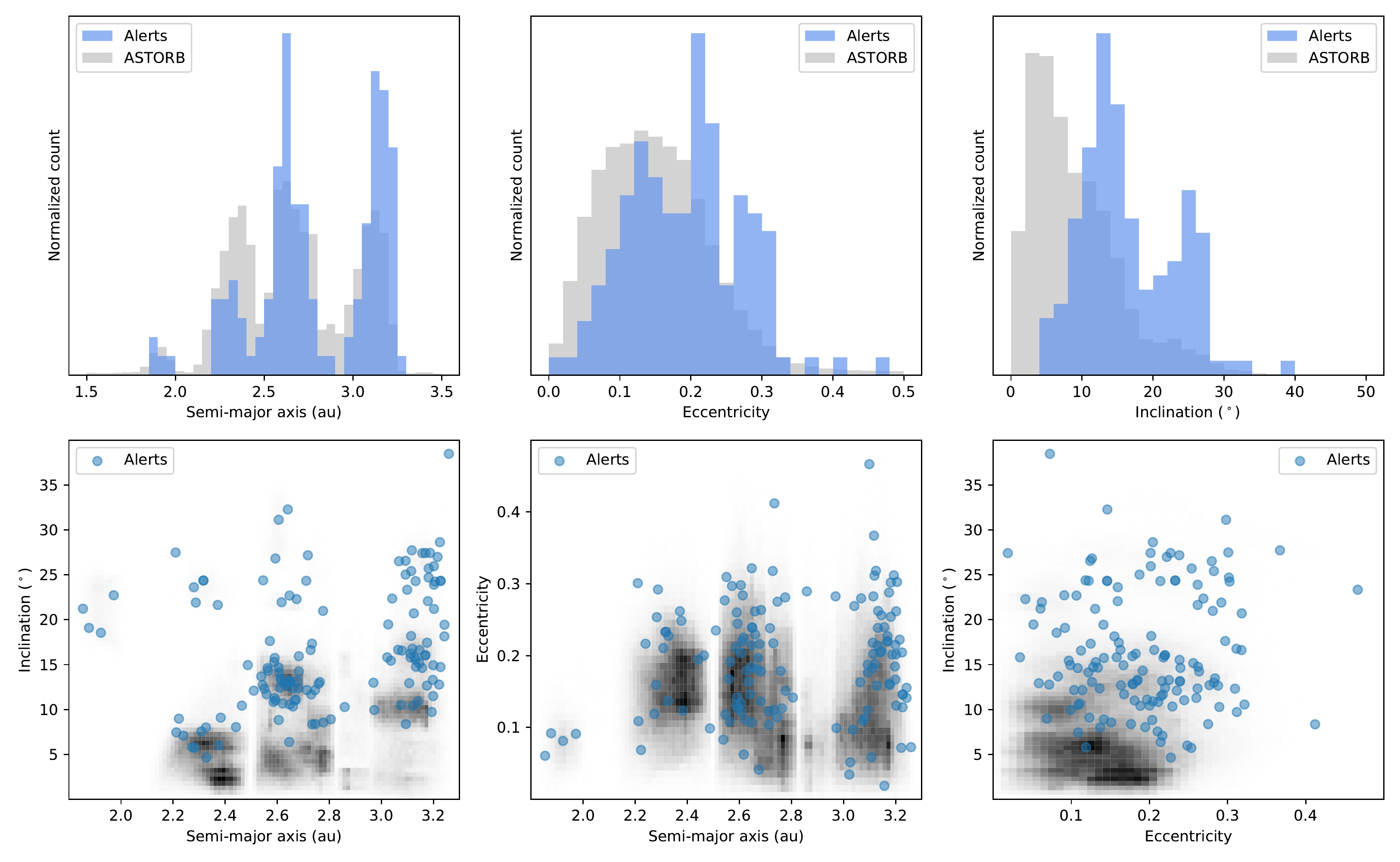}
    \caption{Comparison of the distribution of the detected 
      asteroids with the current census of SSOs as represented by the
      \astorb database. } 
\label{fig:aei}
\end{figure*}

  We use the designations assigned by the MPC to retrieve the orbital 
  elements of the SSOs that led to alerts.
  We compare their distribution in semi-major axis, eccentricity, and
  inclination in \Autoref{fig:aei}.

  As of July 2020, all the recovered \gaia alerts concerned main-belt
  asteroids. Considering that \numb{139} alerts were assigned a designation
  and that the incidence of NEA was expected to be 1--2\%, it is not
  surprising.
  Moreover, the SSO-ST may be biased against NEAs for several reasons.
  First, as NEAs present a higher apparent motion than main-belt asteroids,
  they will smear and move faster outside the transmitted windows.
  Their transits will hence
  contain less observations, which may lead to rejection of the transit
  at the DU454 level.
  Second, their motion combined with the \gaia scanning law may result in fewer overall
  transits, leading to more rejections of NEAs at the DU459 level.
  We can, however, attest the SSO-ST capability of processing NEAs:
  we tested it with known NEAs, and it successfully led to 
	simulated alerts (not released).

  Regarding the distribution of orbital elements, the sample of alerts is 
  different from the current census of asteroids, taken from \astorb.
  By applying 1D Kolmogorov-Smirnov test on semi-major axis, eccentricity, and
  inclination in turn (\Autoref{fig:aei}, top line),
  to test whether the distribution of alerts and 
  \astorb are similar, we find it highly unlikely that the 
  distributions are similar, since the K-S p-values
  are $10^{-4}$, $10^{-9}$, and $10^{-33}$ respectively, well below the 
  typical threshold of 0.2 for similar distributions.

  The alerts are typically located in the outer main-belt,
  and have more inclined and more eccentric orbits than the known population
  (\Autoref{fig:aei}, bottom line).
  There is a weak correlation between semi-major axis and inclination
  (hence alerts are preferentially for high-inclined outer main belt
  asteroids).

  The outer main belt being further, and populated by darker asteroids
  than the inner parts of the belt, its completeness at a given size is
  lower \citep{2013Icar..226..723D}.
  The bias in the current census against high-inclination orbits has already
  been reported by 
  \citet{mahlke2018}, based on deep images obtained by the KiDS survey.
  Here, the distribution of alerts clearly skewed towards
  high inclinations, highlights the strength of \gaia's all-sky survey.
  With the final release of SSO catalogue by \gaia, efforts to debias
  the current census of SSOs will be possible.

\section{Conclusion}

  Since the beginning of \gaia's operations in 2014, a specific pipeline has been
  running daily to process unidentified Solar System Objects observed by \gaia.
  Since 2016, about \numb{1700} calls for observations, called alerts, have been
  released to trigger ground-based follow-up observations of these unidentified objects.
  Among these, \numb{250} recovery observations have been attempted,
  resulting in the detection of 
  \numb{227} of them.
  Their astrometry has been sent to the Minor Planet Center, which assigned a preliminary
  designation to \numb{139} of these objects, including \numb{six} attributed to
  \gaia.
  The orbital distribution of these alerts confirms the bias in the current
  census of the asteroid population against highly inclined objects.
  This bias is even more pronounced in the outer asteroid belt.

  The daily processing pipeline is still running and will continue until 
	the end of \gaia's operations, 
  that have been extended to \numb{the end of 2022} by ESA,
	and possibly until end 2024 (approval pending).

\begin{acknowledgements}

  We are indebted to the engineers of the European Space Agency and 
  of the Data Processing Center CNES in Toulouse, France, 
  who made possible this mission and these observations.
  We thank CNRS-INSU (France) and the scientific council of the \gaia 
  observing service ANO1 for the yearly funding allowing us to 
  observe periodically at Haute-Provence Observatory (OHP).

  This work presents results from the European Space Agency (ESA) space mission 
  \gaia. \gaia data are being processed by the Gaia Data Processing and Analysis Consortium (DPAC). 
  Funding for the DPAC is provided by national institutions, in 
  particular the institutions participating in the Gaia MultiLateral Agreement (MLA). 
  The Gaia mission website is \url{https://www.cosmos.esa.int/gaia}. 
  The Gaia archive website is \url{https://archives.esac.esa.int/gaia}. 

  RAM acknowledges support from CONICYT/FONDECYT Grant Nr. 1190038 
  and from the Chilean Centro de Excelencia en Astrofisica y Tecnologia as Afines (CATA) BASAL PFB/06. 
  We are grateful for the continuous support of the Chilean National Time Allocation 
  Committee under programs CN2019A-3, CN2019B-16, and CN2020A-20.

  We thank all the observers participating in this program. 

  This work has financially been supported by Agenzia Spaziale Italiana (ASI)
  through contracts I/037/08/0, I/058/10/0, 2014-025-R.0.

  The observations at Abastumani were supported by the Shota Rustaveli National Science Foundation,
  Grant RF-18-1193.

\end{acknowledgements}

\bibliographystyle{aa} 
\bibliography{references} 

\begin{thebibliography}{34}
\expandafter\ifx\csname natexlab\endcsname\relax\def\natexlab#1{#1}\fi

\bibitem[{{Bailen} {et~al.}(2020)}]{Bailen20}
{Bailen}, M. {et~al.} 2020, in 51st Lunar and Planetary Science Conference,
  Vol.~51, 2078

\bibitem[{{Bancelin} {et~al.}(2012){Bancelin}, {Hestroffer}, \&
  {Thuillot}}]{2012PSS...73...21B}
{Bancelin}, D., {Hestroffer}, D., \& {Thuillot}, W. 2012, \planss, 73, 21

\bibitem[{Berthier {et~al.}(2016)Berthier, Carry, Vachier, Eggl, \&
  Santerne}]{2016-MNRAS-458-Berthier}
Berthier, J., Carry, B., Vachier, F., Eggl, S., \& Santerne, A. 2016, \mnras,
  458, 3394

\bibitem[{Berthier {et~al.}(2006)Berthier, Vachier, Thuillot, Fernique,
  Ochsenbein, Genova, Lainey, \& Arlot}]{2006-ASPC-351-Berthier}
Berthier, J., Vachier, F., Thuillot, W., {et~al.} 2006, in Astronomical Society
  of the Pacific Conference Series, Vol. 351, Astronomical Data Analysis
  Software and Systems XV, ed. C.~{Gabriel}, C.~{Arviset}, D.~{Ponz}, \&
  S.~{Enrique}, 367

\bibitem[{{Bonnarel} {et~al.}(2000){Bonnarel}, {Fernique}, {Bienaym{\'e}},
  {Egret}, {Genova}, {Louys}, {Ochsenbein}, {Wenger}, \&
  {Bartlett}}]{2000AAS..143...33B}
{Bonnarel}, F., {Fernique}, P., {Bienaym{\'e}}, O., {et~al.} 2000, \aaps, 143,
  33

\bibitem[{{Bouquillon} {et~al.}(2014)}]{Bouquillon14}
{Bouquillon}, S. {et~al.} 2014, in Proceedings of the SPIE, Vol. 9152, 03B

\bibitem[{{Bowell}(2014)}]{Bowell14}
{Bowell}, E. 2014, vizier.u-strasbg.fr\//viz-bin\//VizieR-2

\bibitem[{Brown {et~al.}(2016)Brown, Vallenari, Prusti, de~Bruijne, Mignard,
  Drimmel, Babusiaux, Bailer-Jones, Bastian, \& et~al.}]{2016AA...595A...2G}
Brown, A. G.~A., Vallenari, A., Prusti, T., {et~al.} 2016, \aap, 595, A2

\bibitem[{{Carry}(2014)}]{2014-GFS-Carry}
{Carry}, B. 2014, in Gaia-FUN-SSO workshop 2014

\bibitem[{{Carry} {et~al.}(2019)}]{Carry19}
{Carry}, B. {et~al.} 2019, in EPSC-DPS Joint Meeting 2019, Vol.~13, 1409--1,
  2019

\bibitem[{{Clementini} {et~al.}(2016){Clementini}, {Ripepi}, {Leccia},
  {Mowlavi}, {Lecoeur-Taibi}, {Marconi}, {Szabados}, {Eyer}, {Guy},
  {Rimoldini}, {Jevardat de Fombelle}, {Holl}, {Busso}, {Charnas}, {Cuypers},
  {De Angeli}, {De Ridder}, {Debosscher}, {Evans}, {Klagyivik}, {Musella},
  {Nienartowicz}, {Ord{\'o}{\~n}ez}, {Regibo}, {Riello}, {Sarro}, \&
  {S{\"u}veges}}]{2016AA...595A.133C}
{Clementini}, G., {Ripepi}, V., {Leccia}, S., {et~al.} 2016, \aap, 595, A133

\bibitem[{{Cropper} {et~al.}(2018){Cropper}, {Katz}, {Sartoretti}, {Prusti},
  {de Bruijne}, {Chassat}, {Charvet}, {Boyadjian}, {Perryman}, {Sarri}, {Gare},
  {Erdmann}, {Munari}, {Zwitter}, {Wilkinson}, {Arenou}, {Vallenari},
  {G{\'o}mez}, {Panuzzo}, {Seabroke}, {Allende Prieto}, {Benson}, {Marchal},
  {Huckle}, {Smith}, {Dolding}, {Jan{\ss}en}, {Viala}, {Blomme}, {Baker},
  {Boudreault}, {Crifo}, {Soubiran}, {Fr{\'e}mat}, {Jasniewicz}, {Guerrier},
  {Guy}, {Turon}, {Jean-Antoine-Piccolo}, {Th{\'e}venin}, {David}, {Gosset}, \&
  {Damerdji}}]{2018AA...616A...5C}
{Cropper}, M., {Katz}, D., {Sartoretti}, P., {et~al.} 2018, \aap, 616, A5

\bibitem[{{DeMeo} \& {Carry}(2013)}]{2013Icar..226..723D}
{DeMeo}, F.~E. \& {Carry}, B. 2013, \icarus, 226, 723

\bibitem[{{Fabricius} {et~al.}(2016){Fabricius}, {Bastian}, {Portell},
  {Casta{\~n}eda}, {Davidson}, {Hambly}, {Clotet}, {Biermann}, {Mora},
  {Busonero}, {Riva}, {Brown}, {Smart}, {Lammers}, {Torra}, {Drimmel},
  {Gracia}, {L{\"o}ffler}, {Spagna}, {Lindegren}, {Klioner}, {Andrei}, {Bach},
  {Bramante}, {Br{\"u}semeister}, {Busso}, {Carrasco}, {Gai}, {Garralda},
  {Gonz{\'a}lez-Vidal}, {Guerra}, {Hauser}, {Jordan}, {Jordi}, {Lenhardt},
  {Mignard}, {Messineo}, {Mulone}, {Serraller}, {Stampa}, {Tanga}, {van
  Elteren}, {van Reeven}, {Voss}, {Abbas}, {Allasia}, {Altmann}, {Anton},
  {Barache}, {Becciani}, {Berthier}, {Bianchi}, {Bombrun}, {Bouquillon},
  {Bourda}, {Bucciarelli}, {Butkevich}, {Buzzi}, {Cancelliere}, {Carlucci},
  {Charlot}, {Collins}, {Comoretto}, {Cross}, {Crosta}, {de Felice}, {Fienga},
  {Figueras}, {Fraile}, {Geyer}, {Hernandez}, {Hobbs}, {Hofmann}, {Liao},
  {Licata}, {Martino}, {McMillan}, {Michalik}, {Morbidelli}, {Parsons},
  {Pecoraro}, {Ramos-Lerate}, {Sarasso}, {Siddiqui}, {Steele},
  {Steidelm{\"u}ller}, {Taris}, {Vecchiato}, {Abreu}, {Anglada}, {Boudreault},
  {Cropper}, {Holl}, {Cheek}, {Crowley}, {Fleitas}, {Hutton}, {Osinde},
  {Rowell}, {Salguero}, {Utrilla}, {Blagorodnova}, {Soffel}, {Osorio},
  {Vicente}, {Cambras}, \& {Bernstein}}]{2016AA...595A...3F}
{Fabricius}, C., {Bastian}, U., {Portell}, J., {et~al.} 2016, \aap, 595, A3

\bibitem[{{Fedorets} {et~al.}(2018){Fedorets}, {Muinonen}, {Pauwels},
  {Granvik}, {Tanga}, {Virtanen}, {Berthier}, {Carry}, {David}, {Dell'Oro},
  {Mignard}, {Petit}, {Spoto}, \& {Thuillot}}]{2018-AA-620-Fedorets}
{Fedorets}, G., {Muinonen}, K., {Pauwels}, T., {et~al.} 2018, \aap, 620, A101

\bibitem[{Ivezi{\'c} {et~al.}(2001)Ivezi{\'c}, Tabachnik, Rafikov, Lupton,
  Quinn, Hammergren, Eyer, Chu, Armstrong, Fan, Finlator, Geballe, Gunn,
  Hennessy, Knapp, Leggett, Munn, Pier, Rockosi, Schneider, Strauss, Yanny,
  Brinkmann, Csabai, Hindsley, Kent, Lamb, Margon, McKay, Smith, Waddel, York,
  \& the SDSS~Collaboration}]{2001AJ....122.2749I}
Ivezi{\'c}, {\v Z}., Tabachnik, S., Rafikov, R., {et~al.} 2001, \aj, 122, 2749

\bibitem[{{Jordi} {et~al.}(2010){Jordi}, {Gebran}, {Carrasco}, {de Bruijne},
  {Voss}, {Fabricius}, {Knude}, {Vallenari}, {Kohley}, \&
  {Mora}}]{2010AA...523A..48J}
{Jordi}, C., {Gebran}, M., {Carrasco}, J.~M., {et~al.} 2010, \aap, 523, A48

\bibitem[{{Mahlke} {et~al.}(2018){Mahlke}, {Bouy}, {Altieri}, {Verdoes Kleijn},
  {Carry}, {Bertin}, {de Jong}, {Kuijken}, {McFarland}, \&
  {Valentijn}}]{mahlke2018}
{Mahlke}, M., {Bouy}, H., {Altieri}, B., {et~al.} 2018, \aap, 610, A21

\bibitem[{Mignard {et~al.}(2007)Mignard, Cellino, Muinonen, Tanga, Delbo,
  Dell'Oro, Granvik, Hestroffer, Mouret, Thuillot, \&
  Virtanen}]{2007EMP..101...97M}
Mignard, F., Cellino, A., Muinonen, K., {et~al.} 2007, Earth Moon and Planets,
  101, 97

\bibitem[{{Muinonen} {et~al.}(2016){Muinonen}, {Fedorets}, {Pentik{\"a}inen},
  {Pieniluoma}, {Oszkiewicz}, {Granvik}, {Virtanen}, {Tanga}, {Mignard},
  {Berthier}, {Dell`Oro}, {Carry}, \& {Thuillot}}]{2016PSS..123...95M}
{Muinonen}, K., {Fedorets}, G., {Pentik{\"a}inen}, H., {et~al.} 2016, \planss,
  123, 95

\bibitem[{{Orbfit Consortium}(2011)}]{orbfit}
{Orbfit Consortium}. 2011, {OrbFit: Software to Determine Orbits of Asteroids}

\bibitem[{{Perryman} {et~al.}(2001){Perryman}, {de Boer}, {Gilmore}, {H{\o}g},
  {Lattanzi}, {Lindegren}, {Luri}, {Mignard}, {Pace}, \& {de
  Zeeuw}}]{2001AA...369..339P}
{Perryman}, M.~A.~C., {de Boer}, K.~S., {Gilmore}, G., {et~al.} 2001, \aap,
  369, 339

\bibitem[{{Prusti} {et~al.}(2016)}]{Prusti16}
{Prusti}, T. {et~al.} 2016, \aap, 595, A1

\bibitem[{{Riello} {et~al.}(2018){Riello}, {De Angeli}, {Evans}, {Busso},
  {Hambly}, {Davidson}, {Burgess}, {Montegriffo}, {Osborne}, {Kewley},
  {Carrasco}, {Fabricius}, {Jordi}, {Cacciari}, {van Leeuwen}, \&
  {Holland}}]{2018AA...616A...3R}
{Riello}, M., {De Angeli}, F., {Evans}, D.~W., {et~al.} 2018, \aap, 616, A3

\bibitem[{{Rudenko}(2016)}]{Rudenko16}
{Rudenko}, M. 2016, in Asteroids: New Observations, New Models, Proceedings of
  the International Astronomical Union IAU Symposium, Vol. 318, 265--269

\bibitem[{{Simon} {et~al.}(2019){Simon}, {Pavlenko}, {Shugarov}, {Vasylenko},
  {Izviekova}, {Reshetnyk}, {Godunova}, {Bufan}, {Baransky}, {Antonyuk},
  {Baklanov}, {Troianskyi}, {Udovichenko}, \& {Keir}}]{2019CoSka..49..420S}
{Simon}, A., {Pavlenko}, E., {Shugarov}, S., {et~al.} 2019, Contributions of
  the Astronomical Observatory Skalnate Pleso, 49, 420

\bibitem[{{Spoto} {et~al.}(2018)}]{Spoto18}
{Spoto}, F. {et~al.} 2018, \aap, 614, A27

\bibitem[{{Tanga}(2011)}]{2011sssb.confE...2T}
{Tanga}, P. 2011, in Solar System Science Before and After Gaia, 2

\bibitem[{{Tanga} \& {Thuillot}(2010)}]{2010-GFS-WS}
{Tanga}, P. \& {Thuillot}, W., eds. 2010, Proceedings of the Gaia-FUN SSO
  workshops, IMCCE, Paris Observatory

\bibitem[{{Tanga} \& {Thuillot}(2012)}]{2012-GFS-WS}
{Tanga}, P. \& {Thuillot}, W., eds. 2012, Proceedings of the Gaia-FUN SSO
  workshops, IMCCE, Paris Observatory

\bibitem[{{Tanga} \& {Thuillot}(2014)}]{2014-GFS-WS}
{Tanga}, P. \& {Thuillot}, W., eds. 2014, Proceedings of the Gaia-FUN SSO
  workshops, IMCCE, Paris Observatory

\bibitem[{{Tanga} {et~al.}(2016)}]{Tanga16}
{Tanga}, P. {et~al.} 2016, Planet. Space Sc., 123, 87

\bibitem[{{Thuillot} \& {Dennefeld}(2018)}]{Thuillot18}
{Thuillot}, W. \& {Dennefeld}, M. 2018, in SF2A-2018: Proceedings of the Annual
  meeting of the SF2A, ed. P.~{Di Matteo}, F.~{Billebaud}, F.~{Herpin},
  N.~{Lagarde}, J.~{Marquette}, A.~{Robin}, \& O.~{Venot}, 463--465

\bibitem[{{Thuillot} {et~al.}(2015)}]{Thuillot15b}
{Thuillot}, W. {et~al.} 2015, \aap, 583, 59

\end{thebibliography}

\end{document}